\documentclass[11pt,a4paper,onecolumn]{article}

\title{Is There A Higgs Boson With Mass Below 100\,GeV/c$^2$\\In The LEP Data?}
\author{C.K. Bowdery\footnote{E-mail: chris.bowdery@physics.org}
\\Department of Physics, Lancaster University\\Lancaster,
LA1 4YB, UK}
\date{13 January 2003}
\begin{document}

\maketitle
\begin{abstract}
There could be a Higgs boson with a mass less than 100\,GeV/c$^2$
in the LEP data if its decay width and modes are not dominated by $b$ quarks.
At present, the quantum interference between signal
and background processes has not been considered in flavour independent searches
but when it is, the long-sought particle may appear.
\end{abstract}
\section{Introduction}
The search for a Higgs boson is arguably the top priority of particle physics today.
It is also the most tantalizing. On the one hand, direct searches at LEP~II
have set lower limits of 114\,GeV/c$^2$ \cite{LEPlimit} for a Standard Model Higgs
boson and 89.9\,GeV/c$^2$ for the Minimal Supersymmetric Model Higgs
\cite{PDG}. On the other hand, indirect evidence from virtual Higgs effects in 
electroweak fits to LEP, SLC and Tevatron data suggest a Higgs exists with a mass
significantly lower than these bounds. The fits to the $Z$ data imply a mass of about
81\,GeV/c$^2$ \cite{indirect,LEPEWWG} while the latest fit to the
$W$ data using the measured top quark mass gives a Higgs mass of just 26\,GeV/c$^2$
\cite{Renton}.
Of course, the uncertainties are large (30-50\,GeV/c$^2$) in both types of fit but
nevertheless there is a growing feeling that these fits and the direct search result
are an indication of a deviation from the Standard Model.

The assumption is that, for some reason, the Higgs mass
is larger than the central value predicted
by the fits.
However this may not be so. In a companion paper \cite{mine}
we have proposed a scenario with a universal Higgs-fermion
coupling and its application to a Two Higgs Doublet Model (2HDM) of type~II. With some
reasonable parameters, a Higgs boson with a mass compatible with the fit values but
a hadronic branching ratio of between 66\% and 80\% is predicted.
These hadronic jet flavours are equally
$u$ and $c$ (\emph{not} $b$) with the remaining decays
being to invisible neutrinos. 

Although such a model sidesteps the limits imposed by the standard analyses
quoted above, 
the LEP Higgs Working Group has combined flavour independent Higgs search results
from the four LEP collaborations \cite{ALEPH, DELPHI, L3, OPAL} and produced a paper,
with preliminary results \cite{LHWG}, which appears to impose serious constraints.
At face value, figure~2 from that paper suggests that,
with a $Z$-Higgs production cross
section roughly equal to the Standard Model rate and a hadronic branching ratio of
about 80\%, there can be no Higgs boson with a mass below about 110\,GeV/c$^2$.

\section{Flaws in the Flavour Independent Higgs Search}

The flavour independent Higgs search has two significant flaws.
Firstly the decay width of the Higgs boson in the Universal Higgs-Fermion Coupling
model could be much larger than orthodox models and this needs to be taken into
consideration.

Secondly, and more importantly,
no account has been taken of any quantum interference between the signal
and background. The signal process is
$e^+e^-\rightarrow HZ\rightarrow\ $ 4-jets while
the background processes are $e^+e^-\rightarrow ZZ\rightarrow\ $ 4-jets
and $e^+e^-\rightarrow W^+W^-\rightarrow\ $ 4-jets. The interference may be
significant because
the $H$ could have a similar
mass to the $W$ and $Z$ and similar quark branching ratios.
(Of course, the $Z$ and $W$ are spin one while the
$H$ is spin zero.)

The result of such interference could be dips and rises
in the production cross-section
rather than a simple step.
Whether this is what is seen in figure~2 of the LEP 
Higgs Working Group paper is impossible to tell but it is intiguing. Of course
a careful simulation of the interference (if any) will be needed
and we encourage the LEP Higgs Working Group to undertake this analysis
as soon as possible. If a Higgs boson does hide in the LEP data,
presumably charm tagging as a function of the electron-positron
centre-of-mass energy will be required
to dig it out.

\section{Conclusion}
The current flavour independent Higgs search result from LEP cannot currently exclude
a low mass Higgs boson with properties similar to a $Z$ until quantum interference
is simulated between $HZ$ and $ZZ$ production and the effect of
broad Higgs decay widths
are considered. Consequently the central result from the
indirect (electroweak fit) Higgs searches is still viable.
\section*{Acknowledgements}
I would like to thank Bill Murray for drawing my attention to the flavour independent
search paper.

\end{document}